\documentclass[twocolumn,showpacs,preprintnumbers,amsmath,amssymb,prb]{revtex4-1}

\usepackage{graphicx}
\usepackage{dcolumn}
\usepackage{bm}
\usepackage{tabularx}
\usepackage{amsmath}

\begin{document}

\title{Nodeless superconducting gaps in
Ca$_{10}$(Pt$_{4-\delta}$As$_8$)((Fe$_{1-x}$Pt$_{x}$)$_2$As$_2$)$_5$
probed by quasiparticle heat transport}

\author{X. Qiu, L. P. He, X. C. Hong, Z. Zhang, J. Pan, X. P. Shen, D. L. Feng, and S. Y. Li$^{*}$}

\affiliation{State Key Laboratory of Surface Physics, Department of Physics, and Laboratory of Advanced Materials,
Fudan University, Shanghai 200433, P. R. China
\\}

\date{\today}

\begin{abstract}
The in-plane thermal conductivity of iron-based superconductor
Ca$_{10}$(Pt$_{4-\delta}$As$_8$)((Fe$_{1-x}$Pt$_{x}$)$_2$As$_2$)$_5$ single crystal (``10-4-8", $T_c$ = 22 K)
was measured down to 80 mK. In zero field, the residual linear term $\kappa_0/T$
is negligible, suggesting nodeless superconducting gaps in this multiband compound.
In magnetic fields, $\kappa_0/T$ increases rapidly,
which mimics those of multiband superconductor
NbSe$_2$ and LuNi$_2$B$_2$C with highly anisotropic gap. Such a field dependence of $\kappa_0/T$ is an evidence for multiple
superconducting gaps with quite different magnitudes or highly anisotropic gap. Comparing with the London penetration depth
results of Ca$_{10}$(Pt$_3$As$_8$)((Fe$_{1-x}$Pt$_{x}$)$_2$As$_2$)$_5$ (``10-3-8") compound, the 10-4-8 and 10-3-8 compounds may have
similar superconducting gap structure.

\end{abstract}


\maketitle

\section{Introduction}

To understand the electronic pairing mechanism of a superconductor,
it is very important to know the symmetry and structure of its superconducting gap.
For the iron-based high-temperature superconductors, there are many families,
such as LaO$_{1-x}$F$_{x}$FeAs (``1111''),\cite{LaOFFeAs} Ba$_{1-x}$K$_{x}$Fe$_{2}$As$_{2}$ (``122''),\cite{BaKFeAs} NaFe$_{1-x}$Co$_x$As (``111''),\cite{NaFeCoAs}
and FeSe$_{x}$Te$_{1-x}$ (``11'').\cite{FeSeTe}
The most notable character of these families is the multiple
Fermi surfaces, which may be the reason why their superconducting gap structure
is so complicated.\cite{Hirschfeld,FaWang}

Different from other families, the new compounds
Ca$_{10}$(Pt$_3$As$_8$)((Fe$_{1-x}$Pt$_x$)$_2$As$_2$)$_5$ (``10-3-8")
and
Ca$_{10}$(Pt$_{4}$As$_8$)((Fe$_{1-x}$Pt$_x$)$_2$As$_2$)$_5$ (``10-4-8") consist of
semiconducting [Pt$_{3}$As$_{8}$] layers or
metallic [Pt$_{4}$As$_{8}$] layers sandwiched between
[Fe$_{10}$As$_{10}$] layers, and show superconductivity with maximal $T_c \sim $ 15 and 38 K, respectively.\cite{Catrin,PNAS,JPSJ}
The metallic [Pt$_{4}$As$_{8}$] layers
lead to stronger FeAs interlayer coupling in 10-4-8 compound, thus higher $T_c$ as compared to the 10-3-8 compound.\cite{PNAS}
The upper critical field of both
10-3-8 and 10-4-8 compounds show strong anisotropy.\cite{XHChen,Mun}
For the 10-3-8 compound, the London penetration depth exhibits power-law variation,
which suggests nodeless superconducting gap.\cite{Cho}
For the 10-4-8 compound, the angle-resolved photoemission spectroscopy (ARPES) experiments have revealed a multiband electronic
structure,\cite{DLFeng,Thirupathaiah} but so far there is still no any investigation of its superconducting gap structure.
As the 10-4-8 has a much higher $T_c$ than the 10-3-8 compound,
it will be interesting to study its superconducting gap structure and compare with the 10-3-8 compound.

Low-temperature thermal conductivity measurement is a bulk technique
to study the superconducting gap structure.\cite{HShakeripour}
According to the magnitude of residual linear term
$\kappa_0/T$ in zero field, one may judge whether there exist
gap nodes or not. The field dependence of $\kappa_0/T$ can give further
information on nodal gap, gap anisotropy, or multiple gaps.\cite{HShakeripour}

In this paper, we measure the thermal conductivity of
Ca$_{10}$(Pt$_{4-\delta}$As$_8$)((Fe$_{1-x}$Pt$_{x}$)$_2$As$_2$)$_5$ ($T_c$ = 22 K)
single crystal down to 80 mk. A negligible residual linear term $\kappa_0/T$ is
found in zero magnetic field.
The field dependence of $\kappa_0/T$ is very similar to those in
multigap $s$-wave superconductor NbSe$_2$ and LuNi$_2$B$_2$C with highly anisotropic gap.
Our data strongly suggest
Ca$_{10}$(Pt$_{4-\delta}$As$_8$)((Fe$_{1-x}$Pt$_{x}$)$_2$As$_2$)$_5$ has nodeless superconducting gaps.
The magnitudes of these gaps could be quite different, or some gap may be anisotropic.

\section{Experiment}

Single crystals of Ca$_{10}$(Pt$_{4-\delta}$As$_8$)((Fe$_{1-x}$Pt$_{x}$)$_2$As$_2$)$_5$ ($T_c$ = 22 K)
were grown by the flux method. \cite{DLFeng}
The composition of the sample was determined as
Ca:Fe:Pt:As = 2:1.73:0.79:3.39
by wavelength-dispersive spectroscopy (WDS), utilizing an electron probe microanalyzer (Shimadzu EPMA-1720).
This doping level is close to the $T_c$ = 26 K sample with Ca:Fe:Pt:As = 2:1.8:0.9:3.5 in Ref. 8, which has the
chemical formula
Ca$_{10}$(Pt$_{4-\delta}$As$_8$)((Fe$_{0.97}$Pt$_{0.03}$)$_2$As$_2$)$_5$ ($\delta$ = 0.26) determined by
single crystal structure refinement.
The dc magnetization was measured at $H$ = 20 Oe, with zero-field cooling process, using a SQUID
(MPMS, Quantum Design).

The sample was cleaved to a rectangular shape with dimensions of
1.5 $\times$ 0.74 mm$^2$ in the $ab$ plane and $\sim$25 $\mu$m along the $c$ axis.
Contacts were made directly on the fresh sample surfaces with silver paint, which were used
for both resistivity and thermal conductivity measurements.
In-plane thermal conductivity was measured in a dilution
refrigerator using a standard four-wire steady-state method with two
RuO$_2$ chip thermometers, calibrated $in$ $situ$ against a reference
RuO$_2$ thermometer. Magnetic fields are applied along the $c$ axis. To ensure a
homogeneous field distribution in the samples, all fields are
applied at a temperature above $T_c$.

\section{Results and Discussion}

\begin{figure}
\includegraphics[clip,width=7cm]{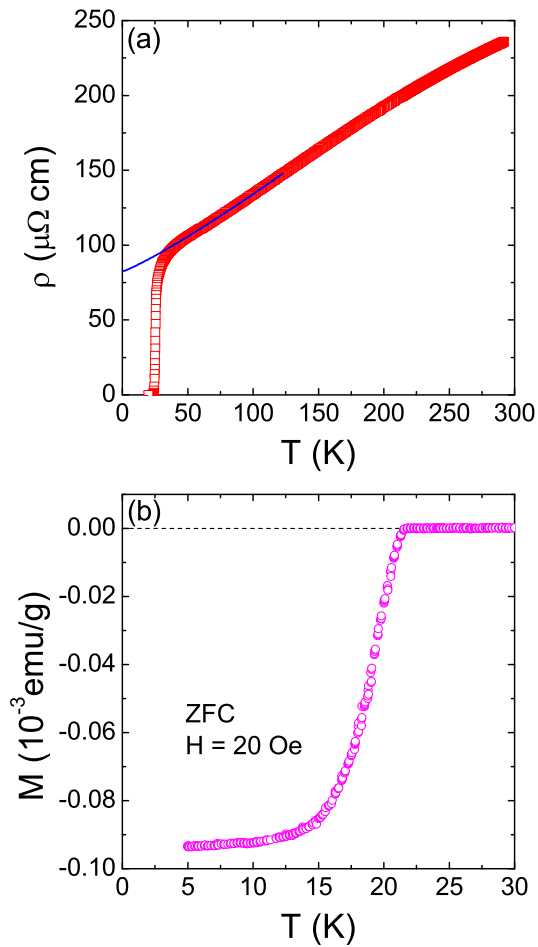}
\caption{(Color online)
(a) In-plane resistivity $\rho(T)$ of
Ca$_{10}$(Pt$_{4-\delta}$As$_8$)((Fe$_{1-x}$Pt$_{x}$)$_2$As$_2$)$_5$
single crystal in zero field. The solid line is a fit of the data between 50 and 125 K
to $\rho = \rho_0 +AT^\alpha$.
(b) Low-temperature magnetization of
Ca$_{10}$(Pt$_{4-\delta}$As$_8$)((Fe$_{1-x}$Pt$_{x}$)$_2$As$_2$)$_5$
single crystal measured in $H$ = 20 Oe, with the zero-field-cooled
(ZFC) process.}
\end{figure}

\begin{figure}
\includegraphics[clip,width=6.5cm]{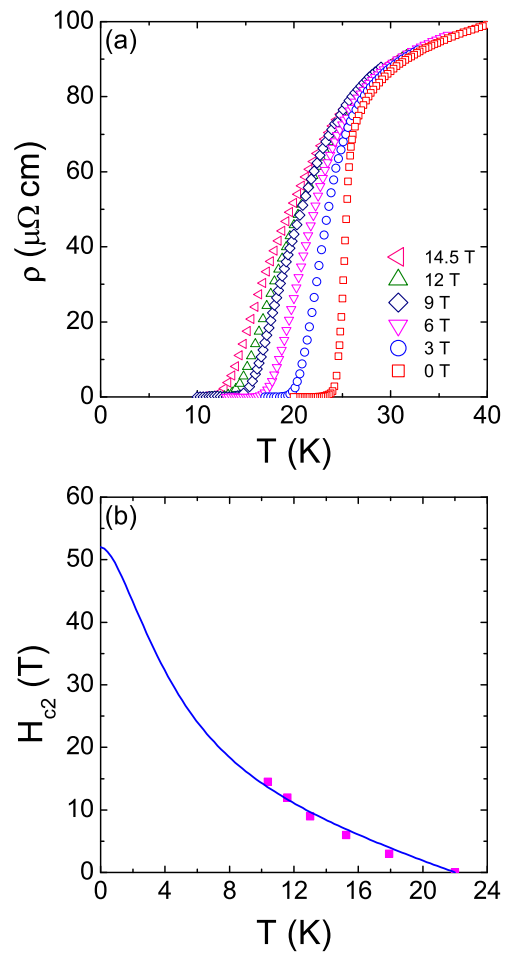}
\caption{(Color online)
(a) Low-temperature resistivity of Ca$_{10}$(Pt$_{4-\delta}$As$_8$)((Fe$_{1-x}$Pt$_{x}$)$_2$As$_2$)$_5$
single crystal in magnetic fields ($H$ = 0, 3, 6, 9, 12, and 14.5 T) along the $c$ axis.
(b) Temperature dependence of the upper critical field $H_{c2}(T)$. The solid line is a fit to the two-band model,\cite{AGurevich,Mun} which points to $H_{c2}$(0) $\approx$ 52 T.}
\end{figure}

Figure 1(a) shows the in-plane resistivity of
Ca$_{10}$(Pt$_{4-\delta}$As$_8$)((Fe$_{1-x}$Pt$_{x}$)$_2$As$_2$)$_5$
single crystal in zero field. Defined by
$\rho$ = 0, $T_c$ = 22.2 K is obtained.
The solid line is a fit of the data between 50 and 125 K
to $\rho = \rho_0 +AT^\alpha$, which gives
residual resistivity $\rho_0$ = 82.5 $\mu \Omega$ cm and $\alpha = 1.15$.
The dc magnetization is shown in Fig. 1(b), and a slightly lower $T_c$ = 21.7 K is found.
Blow we take $T_c$ = 22 K.
This value is lower than the $T_c$ = 38 K at optimal doping. Since the phase diagram,
$T_c$ vs $x$($\delta$), of 10-4-8 system has not been well studied,
it is not sure that our sample is underdoped or overdoped.

Figure 2(a) shows the low-temperature resistivity of
Ca$_{10}$(Pt$_{4-\delta}$As$_8$)((Fe$_{1-x}$Pt$_{x}$)$_2$As$_2$)$_5$
single crystal in magnetic fields up to 14.5 T.
The superconducting transition becomes broader and the $T_c$ decreases with increasing field.
The temperature
dependence of upper critical field $H_{c2}(T)$, defined by $\rho=0$ in Fig. 2(a), is plotted in Fig. 2(b).
To estimate the zero-temperature limit of $H_{c2}$, one usually fits the curve according to the
Werthamer-Helfand-Hohenberg (WHH) theory.\cite{WHH} However, our $H_{c2}(T)$ with upward curvature apparently
can not be fitted well by WHH formula. As explained in Ref. 11, the underlying reason is that the WHH theory is for superconductors with single band,
while the iron-based superconductors have multiple bands. In Ref. 11, the $H_{c2}^{||c}(T)$ curve of
Ca$_{10}$(Pt$_{4-\delta}$As$_8$)((Fe$_{0.97}$Pt$_{0.03}$)$_2$As$_2$)$_5$ ($T_c$ = 26 K) sample was fitted by the two-band model,\cite{AGurevich}
giving $H_{c2}^{||c}(0) \approx $ 90 T.
Taking the same process as in Ref. 11, we also fit the $H_{c2}(T)$ data in Fig. 2(b) with the two-band model,
and get $H_{c2}$(0) = 52 T for our
Ca$_{10}$(Pt$_{4-\delta}$As$_8$)((Fe$_{1-x}$Pt$_{x}$)$_2$As$_2$)$_5$ ($T_c$ = 22 K) sample. Note that a slightly different
$H_{c2}$(0) does not affect our discussion on the field dependence of normalized $\kappa_0/T$ blow.

\begin{figure}
\includegraphics[clip,width=8.5cm]{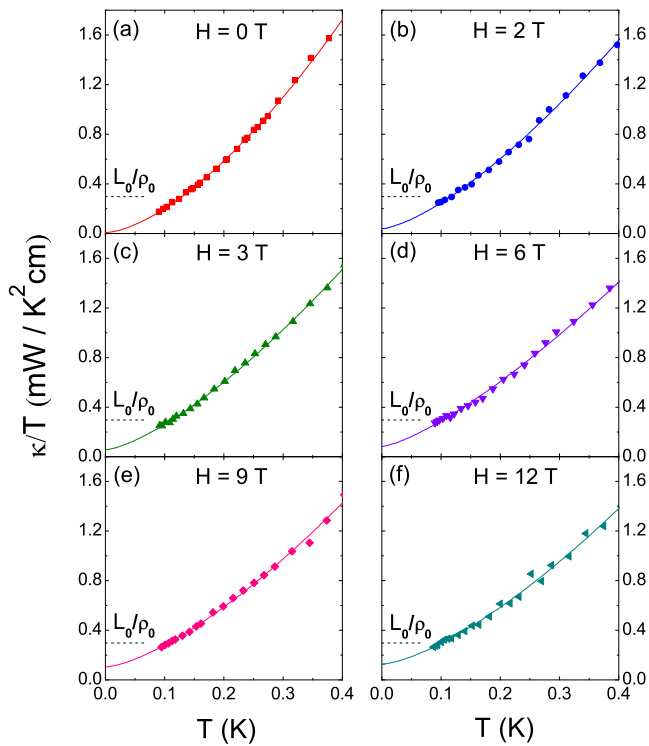}
\caption{(Color online)
Low-temperature thermal conductivity of
Ca$_{10}$(Pt$_{4-\delta}$As$_8$)((Fe$_{1-x}$Pt$_{x}$)$_2$As$_2$)$_5$
with magnetic fields applied along the $c$ axis ($H$ = 0, 2, 3, 6, 9, and 12 T).
The solid lines are $\kappa/T$ = $a + bT^{\alpha-1}$ fits, and the dashed line is the normal-state
Wiedemann-Franz law expectation $L_0$/$\rho_0$.
}
\end{figure}

In Fig. 3, the temperature dependence of the in-plane thermal conductivity for
Ca$_{10}$(Pt$_{4-\delta}$As$_8$)((Fe$_{1-x}$Pt$_{x}$)$_2$As$_2$)$_5$
in $H$ = 0, 2, 3, 6, 9, and 12 T magnetic fields are plotted as $\kappa/T$ vs $T$.
To get the residual linear term $\kappa_0/T$, we fit the curves to
$\kappa/T$ = $a + bT^{\alpha-1}$ blow 0.4 K,
in which the two terms $aT$ and $bT^{\alpha}$ come from
contributions of electrons and phonons, respectively.
\cite{Sutherland,SYLi} The power $\alpha$ of the phonon term is
typically between 2 and 3, due to the
specular reflections of phonons at the sample surfaces. \cite{Sutherland,SYLi}
In zero field, the fitting gives $\kappa_0/T$ = 0.005 $\pm$ 0.013 mW K$^{-2}$ cm$^{-1}$
and $\alpha$ = 2.57 $\pm$ 0.03.
Such a tiny value of $\kappa_0/T$ is within our experimental error bar $\pm$ 0.005 mW K$^{-2}$ cm$^{-1}$. \cite{SYLi}
Therefor it is negligible, comparing to
the normal-state Wiedemann-Franz law expectation $L_0$/$\rho_0$ = 0.297 mW K$^{-2}$ cm$^{-1}$,
with $L_0$ = 2.45 $\times$ 10$^{-8}$ W$\Omega$ K$^{-2}$ and $\rho_0$ = 82.5 $\mu \Omega$ cm.
The absence of $\kappa_0/T$ in zero field means that there are no fermionic quasiparticles to conduct heat as
$T\rightarrow 0$, which provides bulk evidence for nodeless superconducting gaps in
Ca$_{10}$(Pt$_{4-\delta}$As$_8$)((Fe$_{1-x}$Pt$_{x}$)$_2$As$_2$)$_5$, at least in the $ab$ plane.
The data in magnetic fields $H$ = 2, 3, 6, 9, and 12 T are also fitted, as seen in Figs. 3(b)-3(f).
The $\kappa_0/T$ increases significantly with increasing field, although
the maximum applied field $H$ = 12 T is only about 23\% of the $H_{c2}$(0) = 52 T.

\begin{figure}
\includegraphics[clip,width=7.5cm]{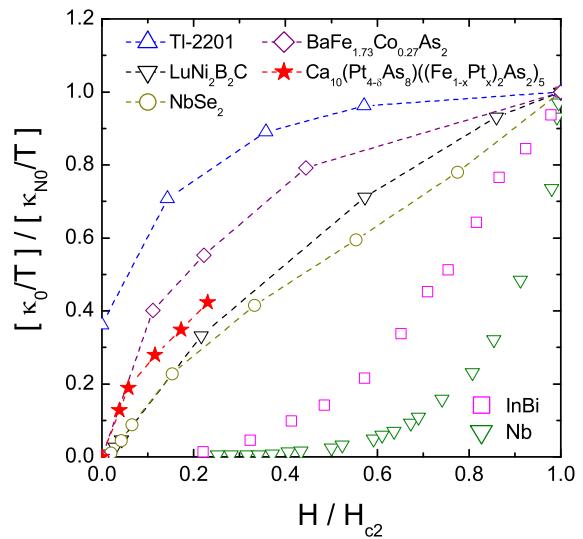}
\caption{(Color online)
Normalized residual linear term
$\kappa_0/T$ of
Ca$_{10}$(Pt$_{4-\delta}$As$_8$)((Fe$_{1-x}$Pt$_{x}$)$_2$As$_2$)$_5$
as a function of $H/H_{c2}$. For comparison, similar data are shown for the clean
$s$-wave superconductor Nb, \cite{Lowell} the dirty $s$-wave
superconducting alloy InBi, \cite{Willis} the multiband $s$-wave
superconductor NbSe$_2$, \cite{Boaknin}
the $s$-wave superconductor LuNi$_2$B$_2$C with highly anisotropic gap,\cite{LuNiBC}
an overdoped $d$-wave cuprate superconductor Tl-2201, \cite{Proust} and the iron-based superconductor BaFe$_{1.73}$Co$_{0.27}$As$_2$.\cite{BaFeCoAs}
}
\end{figure}

To see the field dependence of $\kappa_0/T$ more clearly, the normalized $\kappa_0/T$ of
Ca$_{10}$(Pt$_{4-\delta}$As$_8$)((Fe$_{1-x}$Pt$_{x}$)$_2$As$_2$)$_5$
as a function of $H/H_{c2}$ is plotted in Fig. 4.
Similar data are shown for the clean
$s$-wave superconductor Nb, \cite{Lowell} the dirty $s$-wave
superconducting alloy InBi, \cite{Willis} the multiband $s$-wave
superconductor NbSe$_2$, \cite{Boaknin}
the $s$-wave superconductor LuNi$_2$B$_2$C with highly anisotropic gap,\cite{LuNiBC}
an overdoped $d$-wave cuprate superconductor Tl$_2$Ba$_2$CuO$_{6+\delta}$ (Tl-2201), \cite{Proust} and the iron-based superconductor BaFe$_{1.73}$Co$_{0.27}$As$_2$.\cite{BaFeCoAs}
For clean $s$-wave superconductor with a single isotropic gap, $\kappa_0/T$ is expected to
grow exponentially with increasing $H$, as found in Nb.\cite{Lowell}
In a $d$-wave superconductor, $\kappa_0/T$ increases roughly proportional to $\sqrt{H}$
at low field due to the Volovik effect,\cite{Volovik} as found in Tl-2201.\cite{Proust}

From Fig. 4, the normalized $\kappa_0/T$ of our 10-4-8 compound starts from a negligible value at zero field,
then increases very rapidly with increasing field. This behavior clearly mimics those of NbSe$_2$ and LuNi$_2$B$_2$C. \cite{Boaknin,LuNiBC}
For the multiband $s$-wave superconductor NbSe$_2$, the gap on the $\Gamma$ band is approximately one third of the gap on the other
two Fermi surfaces and magnetic field first suppresses the superconductivity on the Fermi surface with smaller gap.\cite{Boaknin}
For the $s$-wave superconductor LuNi$_2$B$_2$C with highly anisotropic gap, the gap minimum $\Delta_{min}$ is at least 10 times smaller
than the gap maximum, $\Delta_{min} \leq \Delta_0/10$.\cite{LuNiBC}
The nearly identical field dependence of normalized $\kappa_0/T$ between NbSe$_2$ and LuNi$_2$B$_2$C indicates that bulk thermal conductivity measurement
can not distinguish these two kinds of superconducting gap structures.
Nevertheless, the field dependence of $\kappa_0/T$ suggests that the nodeless superconducting gaps in multiband 10-4-8 compound may have
quite different magnitudes, or some gap could be anisotropic. Note that similar field dependence of $\kappa_0/T$ was also observed in
iron-based superconductors BaFe$_{1.73}$Co$_{0.27}$As$_2$ and FeSe$_x$.\cite{BaFeCoAs,FeSe}

In a theoretical calculation of $\kappa_0(H)/T$ with unequal size of isotropic $s_{\pm}$-wave gaps,
the shape of $\kappa_0(H)/T$ changes systematically with the gap size ratio $\Delta _{S}/\Delta _{L}$.\cite{Bang}
In case of isotropic $s$-wave gaps with unequal size, the ratio $\Delta _{S}/\Delta _{L} \approx 1/4$ is estimated for our 10-4-8 compound,
by comparing with the theoretical curves. However, we can not rule out that some gap may be anisotropic.
In fact, the robust power-law variation of London penetration depth observed in 10-3-8 compound was interpreted as a multigap behavior,
and the anisotropy of some superconducting gaps may increase towards the edges of the superconducting dome. \cite{Cho}
In the sense that both thermal conductivity and London penetration depth measurements are bulk probe of the low-energy quasiparticles, the 10-4-8 and 10-3-8 compounds may have similar superconducting gap structure.

\section{Summary}

In summary, we have measured the thermal conductivity of
Ca$_{10}$(Pt$_{4-\delta}$As$_8$)((Fe$_{1-x}$Pt$_{x}$)$_2$As$_2$)$_5$
single crystal down to 80 mK.
The absence of $\kappa_0/T$ in zero field gives strong evidence for nodeless superconducting gaps in such a multiband compound.
The rapid field dependence of $\kappa_0/T$ suggests multiple
superconducting gaps with quite different magnitudes or highly anisotropic gap,
which may be similar to that of Ca$_{10}$(Pt$_3$As$_8$)((Fe$_{1-x}$Pt$_{x}$)$_2$As$_2$)$_5$ compound.

\begin{center}
{\bf ACKNOWLEDGEMENTS}
\end{center}

This work is supported by the Natural Science Foundation
of China, the Ministry of Science and Technology of China
(National Basic Research Program
No. 2012CB821402), and the Program for Professor of Special
Appointment (Eastern Scholar) at Shanghai Institutions of
Higher Learning.

$^*$ E-mail: shiyan$\_$li@fudan.edu.cn


\begin{thebibliography}{99}

\bibitem{LaOFFeAs} Y. Kamihara, T. Watanabe, M. Hirano, and H. Hosono, J. Am. Chem. Soc. {\bf 130}, 3296 (2008).
\bibitem{BaKFeAs} M. Rotter, M. Tegel, and D. Johrendt, Phys. Rev. Lett. {\bf 101}, 107006 (2008).
\bibitem{NaFeCoAs} D. R. Parker, M. J. P. Smith, T. Lancaster, A. J. Steele, I. Franke, P. J. Baker, F. L. Pratt,
    M. J. Pitcher, S. J. Blundell, and S. J. Clarke, Phys. Rev. Lett. {\bf 104}, 057007 (2010).
\bibitem{FeSeTe} B. C. Sales, A. S. Sefat, M. A. McGuire, R. Y. Jin, D. Mandrus, and Y. Mozharivskyj, Phys. Rev. B {\bf 79}, 094521 (2009).
\bibitem{Hirschfeld} P. J. Hirschfeld, M. M. Korshunov, and I. I. Mazin, Rep. Prog. Phys. {\bf 74}, 124508 (2011).
\bibitem{FaWang} Fa Wang and D.-H. Lee, Science {\bf 332}, 200 (2011).
\bibitem{Catrin} C. L\"{o}hnert, T. St\"{u}rzer, M. Tegel, R. Frankovsky, G. Friederichs, and D. Johrendt,
    Angew. Chem. Int. Ed. {\bf 50}, 9195 (2011).
\bibitem{PNAS}  N. Ni, J. M. Allred, B. C. Chan, and R. J. Cava,
    Proc. Natl. Acad. Sci. USA {\bf 108}, E1019 (2011).
\bibitem{JPSJ} S. Kakiya, K. Kudo, Y. Nishikubo, K. Oku, E. Nishibori, H. Sawa,
    T. Yamamoto, T. Nozaka, and M. Nohara, J. Phys. Soc. Jpn. {\bf 80}, 093704 (2011).
\bibitem{XHChen} Z. J. Xiang, X. G. Luo, J. J. Ying, X. F. Wang, Y. J. Yan, A. F. Wang, P. Cheng, G. J. Ye, and X. H. Chen, Phys. Rev. B {\bf 85}, 224527 (2012).
\bibitem{Mun} E. Mun, N. Ni, J. M. Allred, R. J. Cava, O. Ayala, R. D. McDonald,
    N. Harrison, and V. S. Zapf, Phys. Rev. B {\bf 85}, 100502(R) (2012).
\bibitem{Cho} K. Cho, M. A. Tanatar, H. Kim, W. E. Straszheim, N. Ni, R. J. Cava, and R. Prozorov,
    Phys. Rev. B {\bf 85}, 020504(R) (2012).
\bibitem{DLFeng} X. P. Shen, S. D. Chen, Q. Q. Ge, Z. R. Ye, F. Chen, H. C. Xu, S. Y. Tan, X. H. Niu,
    Q. Fan, B. P. Xie, and D. L. Feng,
    Phys. Rev. B {\bf 88}, 115124 (2013).
\bibitem{Thirupathaiah} S. Thirupathaiah, T. St\"{u}rzer, V. B. Zabolotnyy, D. Johrendt, B. B\"{u}chner, and S. V. Borisenko,
    Phys. Rev. B {\bf 88}, 140505(R) (2013).
\bibitem{HShakeripour} H. Shakeripour, C. Petrovic, and L. Taillefer,
    New J. Phys. {\bf 11}, 055065 (2009).
\bibitem{WHH} N. R. Werthamer, E. Helfand, and P. C. Hohemberg, Phys. Rev. {\bf 147}, 295 (1966).
\bibitem{AGurevich} A. Gurevich,
    Phys. Rev. B {\bf 67}, 184515 (2003).
\bibitem{Sutherland} M. Sutherland, D. G. Hawthorn, R. W. Hill, F. Ronning, S. Wakimoto, H. Zhang, C. Proust, E. Boaknin, C. Lupien, L. Taillefer, R. Liang, D. A. Bonn, W. N. Hardy, R. Gagnon, N. E. Hussey, T. Kimura, M. Nohara, and H. Takagi,
    Phys. Rev. B {\bf 67}, 174520 (2003).
\bibitem{SYLi} S. Y. Li, J.-B. Bonnemaison, A. Payeur, P. Fournier, C. H. Wang, X. H. Chen, and L. Taillefer,
    Phys. Rev. B {\bf 77}, 134501 (2008).
\bibitem{Lowell} J. Lowell and J. B. Sousa, J. Low. Temp.
    Phys. {\bf 3}, 65 (1970).
\bibitem{Willis} J. O. Willis and D. M. Ginsberg,
    Phys. Rev. B {\bf 14}, 1916 (1976).
\bibitem{Boaknin} E. Boaknin, M. A. Tanatar, J. Paglione, D. Hawthorn, F. Ronning, R. W. Hill, M. Sutherland, L. Taillefer, J. Sonier, S. M. Hayden, and J. W. Brill,
    Phys. Rev. Lett. {\bf 90}, 117003 (2003).
\bibitem{LuNiBC} E. Boaknin, R. W. Hill, C. Proust, C. Lupien, L. Taillefer, and P. C. Canfield, Phys. Rev. Lett. {\bf 87}, 237001 (2001).
\bibitem{Proust} C. Proust, E. Boaknin, R. W. Hill, L. Taillefer, and A. P. Mackenzie,
    Phys. Rev. Lett. {\bf 89}, 147003 (2002).
\bibitem{BaFeCoAs} J. K. Dong, S. Y. Zhou, T. Y. Guan, X. Qiu, C. Zhang, P. Cheng, L. Fang, H. H. Wen, and S. Y. Li, Phys. Rev. B {\bf 81}, 094520 (2010).
\bibitem{Volovik} G. E. Volovik, JETP Lett. {\bf 58}, 469 (1993).
\bibitem{FeSe} J. K. Dong, T. Y. Guan, S. Y. Zhou, X. Qiu, L. Ding, C. Zhang, U. Patel, Z. L. Xiao, and S. Y. Li,
    Phys. Rev. B {\bf 80}, 024518 (2009).
\bibitem{Bang} Y. Bang, Phys. Rev. Lett. {\bf 104}, 217001 (2010).

\end{thebibliography}
\end{document}